# ANALYSIS OF EINSTEIN'S FORMULA FOR TIME COURSE CHANGE IN NONINERTIAL REFERENCE FRAMES

M.A. Samokhvalov

Tomsk state university
E-mail: samohvalovma@tpu.ru

It is shown in this work that time course change in noninertial reference frames, which were discovered by Einstein, leads to alteration of the formulas for the relativistic mass and the relativistic energy.

In one of the first articles on relativity theory [1] Einstein had deduced the formula for time course change in the noninertial reference frame in uniformly accelerated and rectilinear motion relatively to the inertial reference system:

$$\tau = \frac{\sigma}{\left(1 + \frac{\gamma x}{c^2}\right)}, \quad (1)$$

where: $\sigma$ - time, which had passed in the inertial reference system since noninertial reference system motion starting; $\gamma$ - acceleration of the noninertial reference system in the inertial reference system; $x$ - distance passed by noninertial reference system during the time $\sigma$, $c$ - speed of light in vacuum in the inertial reference system; $\tau$ - time, which had passed in the noninertial reference system since its motion starting. For reasons of convenience formula (1) values designations will be changed in the following way: $\sigma \rightarrow t_0$; $\gamma \rightarrow a_0$; $\tau \rightarrow t$; $x \rightarrow r_0$; $c \rightarrow c_0$. As a result we have:

$$t = \frac{t_0}{\left(1 + \frac{a_0 r_0}{c_0{}^2}\right)}. \quad (2)$$

We transform the formula (2), taking into consideration that $a_0 = \frac{d\vartheta_0}{dt_0} = const$, $r_0 = \int_0^{r_0} dr_0$, $dr_0 = \vartheta_0 dt_0$:

$$t = t_0 \frac{1}{1 + \frac{\int_0^{\vartheta_0} \frac{d\vartheta_0}{dt_0} \vartheta_0 dt_0}{c_0{}^2}} = t_0 \frac{1}{1 + \frac{(\vartheta_0)^2}{2c_0{}^2}}, \quad (3)$$

where: $\vartheta_0$ - actual speed of the noninertial reference system relatively to the inertial one. In the formula (3) without any complications we can substitute $t$ and

$t_0$ with infinitely small time intervals $dt$ and $dt_0$. As a result we have the following formula:

$$dt = dt_0 \frac{1}{1+\frac{a_0 r_0}{c_0^2}} = dt_0 \frac{1}{1+\frac{\vartheta_0^2}{2c_0^2}}. \quad (4)$$

Due to infinite smallness of the time intervals, the formula (4) is valid not only in case of uniform rectilinear motion of the noninertial reference system, but also in case of it's any curvilinear and nonuniform motion (see Einstein's notes in [1]).

In this article мы will analyze implication of the formula (4) for experiments on relativistic mass.

Before we start analyzing corollaries of the formula (4) let's outline the following remark. As we will use differential equations of material body dynamics, then for proper estimation of parameters of body dynamics and kinematics while writing of dynamics equation we should take time, running in the noninertial reference system, which is rigidly connected with the moving body. In this regard, during the time $dt$ material body will go the smaller distance in the inertial reference system, than during the time $dt_0$. Therefore speed $\vartheta$, which is measured with use of timer, that is rigidly connected to material body, should be taken for material body dynamics and kinematics equations. The formula for substitution of speed $\vartheta_0$, which is measured with timer that is motionless in the inertial reference system, with speed $\vartheta$ is the direct corollary the formula (4) and is represented as:

$$\vartheta = \vartheta_0 \left(1+\frac{a_0 r_0}{c_0^2}\right) = \vartheta_0 \left(1+\frac{\vartheta_0^2}{2c_0^2}\right). \quad (5)$$

From these positions we analyze motion of charged particle with charge $e$, mass $m_0$, relativistic mass $m$, linear speed $\vartheta_0$ and $\vartheta$ in with induction $\vec{B}$. Parameters of such motion were widely used for validation of the following formula, which describes relativistic mass changes [2]:

$$m = \frac{m_0}{\left(1-\frac{\vartheta_0^2}{c_0^2}\right)^{\frac{1}{2}}}. \quad (6)$$

For this case, with consideration of (5) and (6), relativistic equation of particle dynamics in the inertial reference system is:

$$m\vartheta^2 \vec{R}^{(-1)} = e\left[\vec{\vartheta}\vec{B}\right]. \quad (7)$$

Experimentally confirmed traditional relativistic equation of dynamics for such case is:

$$\frac{m_0}{\left(1-\frac{\vartheta_0^2}{c_0^2}\right)^{\frac{1}{2}}}\vartheta_0^2\vec{R}^{(-1)}=e\left[\vec{\vartheta}_0\vec{B}\right] \quad . \qquad (8)$$

If one inserts expressions from formulas (5), (6) to the equation (7) there would be following equation of relativistic dynamics of charged particle in transverse magnetic field:

$$\frac{m_0\left(1+\frac{\vartheta_0^2}{2c_0^2}\right)}{\left(1-\frac{\vartheta_0^2}{c_0^2}\right)^{\frac{1}{2}}}\vartheta_0^2\vec{R}^{(-1)}=e\left[\vec{\vartheta}_0\vec{B}\right] \quad . \qquad (9)$$

This equation isn't consistent with results of experiments on relativistic mass. There is only one opportunity to match equation (7) with results of experiments on relativistic mass without serious complications. One needs to assume that relativistic mass is:

$$m=\frac{m_0}{\left(1+\frac{a_0r_0}{c_0^2}\right)\left(1-\frac{\vartheta_0^2}{c_0^2}\right)^{\frac{1}{2}}}=\frac{m_0}{\left(1+\frac{\vartheta_0^2}{2c_0^2}\right)\left(1-\frac{\vartheta_0^2}{c_0^2}\right)^{\frac{1}{2}}} \; . \qquad (10)$$

In this case equation (7) is brought to equation (8) and consistent with experimental data. At that it's more consistent with results of recent experiments [3], [4] on relation between kinetic energy of charged particle and it's momentum at motion in transverse magnetic field. Now take a look at this compliance.

Let's begin with paper [3]. Like in paper [4] instrumentation for measuring of momentum and energy of relativistic electrons from different $\beta$ - sources was used for this work. One may rewrite customary equation of relativistic dynamics (8) in scalar form:

$$p=\frac{m_0}{\left(1-\frac{\vartheta_0^2}{c_0^2}\right)^{\frac{1}{2}}}\vartheta_0=eBR \quad , \qquad (11)$$

where $p$ - electron momentum. Likewise one may insert expressions of $\vartheta$ and $m$ from formulas (5) and (10) and rewrite dynamics equation (7):

$$\frac{m_0}{\left(1+\frac{\vartheta_0^2}{2c_0^2}\right)\left(1-\frac{\vartheta_0^2}{c_0^2}\right)^{\frac{1}{2}}}\vartheta_0\left(1+\frac{\vartheta_0^2}{2c_0^2}\right)=eBR \quad \text{or} \quad \frac{m_0}{\left(1-\frac{\vartheta_0^2}{c_0^2}\right)^{\frac{1}{2}}}\vartheta_0=eBR . \quad (12)$$

Right part of equation (12) for maximal energy electrons from $\beta$ - source $^{204}Tl$ was retrieved experimentally:

$$eBR = 1190 \text{ KeV/c}_0 . \quad (13)$$

At that relation of particle speed to speed of light was retrieved from equation (11) or (12):

$$\frac{\vartheta_0}{c_0} = 0{,}918866 . \quad (14)$$

Then estimation of electron kinetic energy $K$ was carried out with use of already known equivalent formulas:

$$K = \frac{m_0 c_0^2}{\left(1-\frac{\vartheta_0^2}{c_0^2}\right)^{\frac{1}{2}}} - m_0 c_0^2 \quad \text{or} \quad K = \left[\left(pc_0\right)^2 + \left(m_0 c_0^2\right)^2\right]^{\frac{1}{2}} - m_0 c_0^2 \quad (15)$$

It was proved to be equal to $810$ KeV. Meanwhile maximal value of kinetic energy of electrons from $\beta$ - source $^{204}Tl$ is equal to $760$ KeV.

Estimation by classical formula for kinetic energy $K = \frac{p^2}{2m_0}$ carried out in the work [3] had given $1390$ KeV.

In present case quasiclassical formula for kinetic energy is:

$$K = \frac{m\vartheta^2}{2} = \frac{m_0\vartheta_0^2\left(1+\frac{a_0 r_0}{c_0^2}\right)}{2\left(1-\frac{\vartheta_0^2}{c_0^2}\right)^{\frac{1}{2}}} = \frac{m_0\vartheta_0^2\left(1+\frac{\vartheta_0^2}{2c_0^2}\right)}{2\left(1-\frac{\vartheta_0^2}{c_0^2}\right)^{\frac{1}{2}}} . \quad (16)$$

Kinetic energy estimation with use of this formula with consideration of (14) gives $777{,}5284$ KeV, that is significantly more consistent with known maximal energy of electrons from $\beta$ - source $^{204}Tl$ which is equal to $760$ KeV.

Even more consensual are the results of calculation with use of formula (16) with experimental data according to the results of the work [4]. There are no experimental tabular data in that work, so we have taken figure 6 from there, which contains calculated data as graphs and experimental data as squares. Figure 6 from paper [4] is supplemented with a graph (black line), that was drawn according to electrons kinetic energy calculation results using the formula (16) with preliminary estimation of $\frac{\vartheta_0}{c_0}$ ratio through electrons momenta (see figure 1).

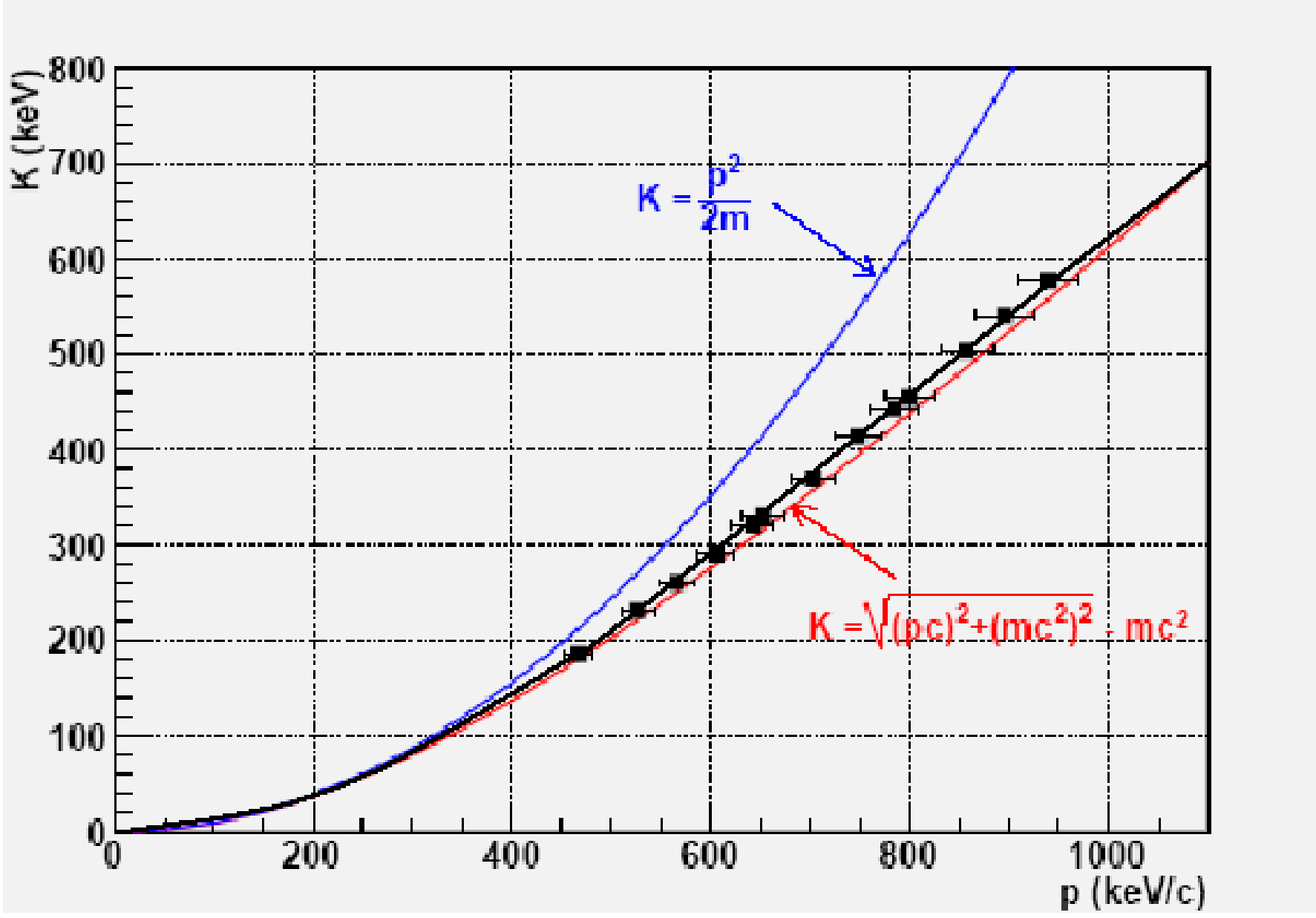

Fig. 1. Figure 6 taken from paper [4]. Results of estimation of electrons kinetic energy (blue and red lines) and experimental values of kinetic energy (black squares) are shown. We have added black line that was drawn using the quasiclassical formula for kinetic energy (16), which considers time course changes in the noninertial reference system related to electrons.

In such a way, noninertial reference system time course change, which was discovered by Einstein, inevitably leads to changing of formulas for relativistic mass and relativistic energy. Although this statement is consistent with already known experimental data it requires further analysis and verification.